# Design of a multichannel 127 °cylindrical spectrometer for sputtered ions

Deyang Yu[a)]

*Institute of Modern Physics, Chinese Academy of Sciences, Lanzhou 730000, China*

[a)]Electronic mail: *d.yu@impcas.ac.cn*

Design details of a 127 ° electrostatic cylindrical spectrometer equipped with a position-sensitive micro-channel plate detector for measuring the sputtered ions in collisions of highly charged ions with solid surface is described. The nonlinear relationship between the point of fall versus the ionic energy, the blurring of the point of fall caused by the divergence of incident angle and the finite entrance aperture, the transform from a position spectrum to an energy spectrum, as well as the influence of the fringing fields are discussed.

## I. INTRODUCTION

A 127 ° cylindrical deflector is often used as an energy analyzer or a nomochromator in many scientific fields owing to its simple structure.[1-5] Since the pioneer work of Hughes, Rojansky and McMillen,[6, 7] the trajectory of a charged particle in a cylindrical electrostatic field has been comprehensively studied. For examples, Hughes and Rojansky pointed out that the receiving slit should be placed at the angle of $\pi/\sqrt{2} \approx 127°$ downstream to the entrance slit in a cylindrical electrostatic field, rather than $\pi/2$ or $\pi$,[6] and therefore the name 127 ° cylindrical deflector or analyzer came. The refocusing properties at $\pi/\sqrt{2}$ were also verified in their following work.[7] Arnow and Jones reanalyzed the trajectories, considered the effects of a finite sized source, and proposed an alternative tandem scheme.[8, 9] Tomková *et al.* emphasized the importance of the relative position and direction of an analyzer to a source, according to their trajectory calculation of a 127 ° secondary electron analyzer.[10] These earlier works concentrate on the trajectory property nearby the circular orbit and lay a solid foundation for its application as an energy analyzer or a nomochromator.[7, 10-19]

Theoretically, the trajectory of a charged particle in an idea cylindrical electrostatic field is governed by the Binet equation[20]

$$h^2 u(\frac{d^2 u}{d\theta^2} + u) = \frac{qk}{m}, \qquad (1)$$

Where $u = 1/r$ is a function of the revolving angle $\theta$, $m$ is the mass, $q$ is the charge state, the constant $k$ represents the central electrostatic field by $\varepsilon(r) = -k/r$, and $h = L/m$ depends on the conserved angular momentum $L$. For any given field $\varepsilon(r) = -k/r$, there exist circular orbits with a kinetic energy of $E_c = q\kappa/2$, and interestingly it is independent of the radius of the orbits. Hereafter we call $q\kappa/2$ *the circular orbit energy*. It is well known that the trajectory $r(\theta)$ is a periodic function with a period of $\sqrt{2}\pi$. The particles with a kinetic energy of the circular orbit energy, starting from a same point at the radius of $r_0$ almost perpendicularly but with small direction divergence will go nearby a circle, and will be refocused to the same radius after revolving an angle of $\pi/\sqrt{2}$ approximately.[6, 8] On the contrary, the particles with other kinetic energies will be separated out of the circular orbit at the revolving angle of $\pi/\sqrt{2}$.

A significant step forward was made by Oshima, Franchy and Ibach.[21, 22] They numerically simulated those trajectories do not close to the circular orbit energy and introduced the multichannel scheme. After that, several multichannel 127 ° cylindrical spectrometers were constructed by employing position-sensitive detectors.[23, 24] Because a broad range of an energy spectrum can be recorded in parallel, when it is employed to analyze an energy spectrum the efficiency can be improved up to two order of magnitude.

We have constructed a 127 ° cylindrical energy spectrometer equipped with a position-sensitive micro-channel plate (MCP) detector, to measure the spectra of the sputtered ions in highly charged ions (HCIs) with surface collisions. It has been already employed to measure energy spectra of sputtered ions in 0.8–1.8 MeV Ar[8+, 9+] with surfaces of beryllium and highly oriented pyrolytic graphite interactions.[25] In this paper, design details including application requirements, trajectory simulations, error estimation and control, engineering realization and spectrum transform are described.

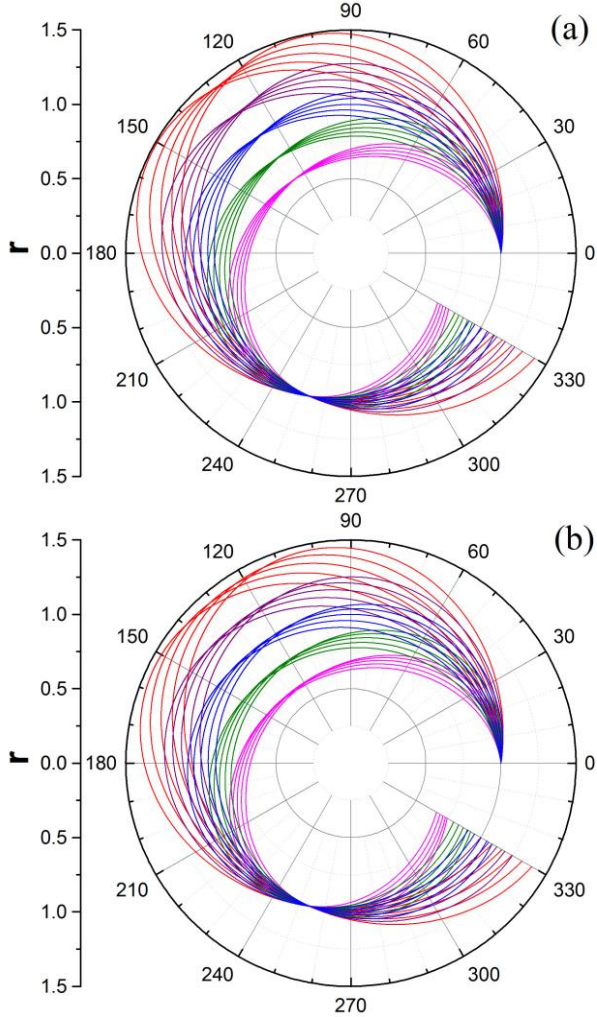

FIG. 1. The trajectories of ions starting from a same point but with different energies and different incident angles. (a) The ions belong to a same bunch have a same angular momentum, and (b) have a same energy. The divergence of the incident angle between two neighboring trajectories is 4°, and the energy of the central trajectory of each bunch is 0.6, 0.8, 1.0, 1.2 and 1.4 time of the circular orbit energy, respectively.

## II. DESIGN

In the present application, an HCI beam, with its diameter of cross section as large as Φ5 mm, is utilized to bombard a solid surface. The reaction products, including electrons, photons, sputtered and scattered atoms and ions go out from the interaction area on the target surface and propagate in a field-free vacuum chamber. The energy distribution of the sputtered ions is interested, which depends on the beam, the target, as well as the incident angle and the observing angle, because it carries important information of the interaction process. We are required to give consideration to both the efficiency of the spectrometer as well as the resolution of energy and observing angle. It should be noted that the interaction area (i.e., the source of the sputtered ions) on the target surface is a finite uniform ellipse rather than a point source. The incident beam is monitored by a real-time beam current density meter,[26] and the absolute differential yields is possible to be deduced by precisely collimating the source area.

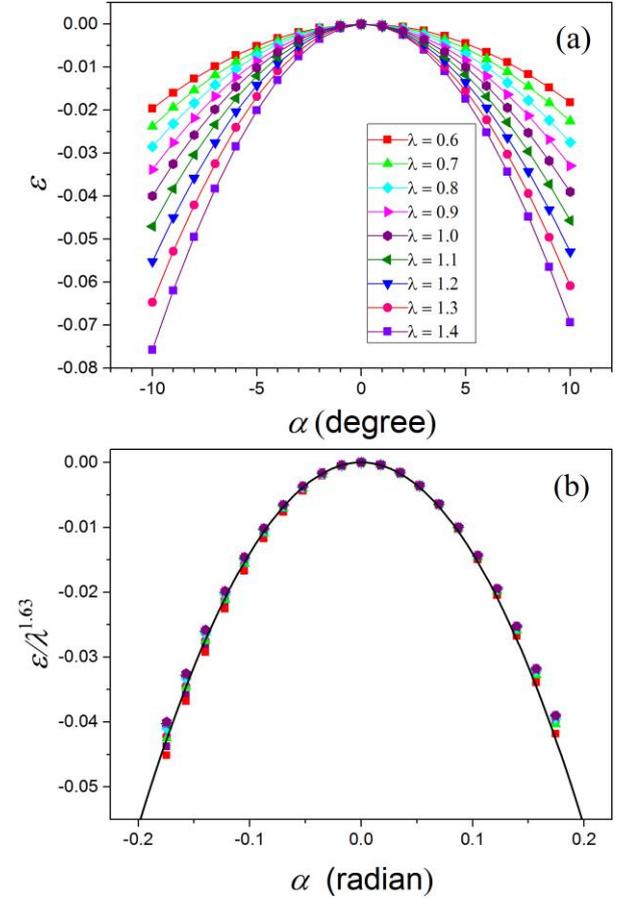

FIG. 2. (a) Blurring of the point of fall $\varepsilon = (r_1' - r_1)/r_0$ at $\pi/\sqrt{2}$ caused by the incident angle $\alpha$ depending on the relative energy $\lambda$. (b) The reduced data of $\varepsilon/\lambda^{1.63}$ versus the incident angle $\alpha$, with a solid parabola of $-1.4\alpha^2$. Note for convenience the unit of angle in (a) and (b) is degree and radian, respectively.

As a starting point, calculation from Eq. (1) of the idea field is made to obtain some general properties of the trajectory clusters, from which the design criteria are drawn. Without loss of generality, assuming the particle starts from $\theta=0$ and hence $u(0)=1/r(0)$.

Define the normal incident angle is perpendicular to the radius, and represent the tilt angle deviating from the normal incident by $\alpha$, hence $h = v\cos\alpha$ and $u'(0) = \tan\alpha / r(0)$. The calculation has been dimensionlessly realized in a C++ program, and a direct solving of the Newtonian equation of motion has also been programmed to verify algorithms. Although the fringing fields do influence the trajectories, it will be introduced by employing the SIMION software[27] and discussed later.

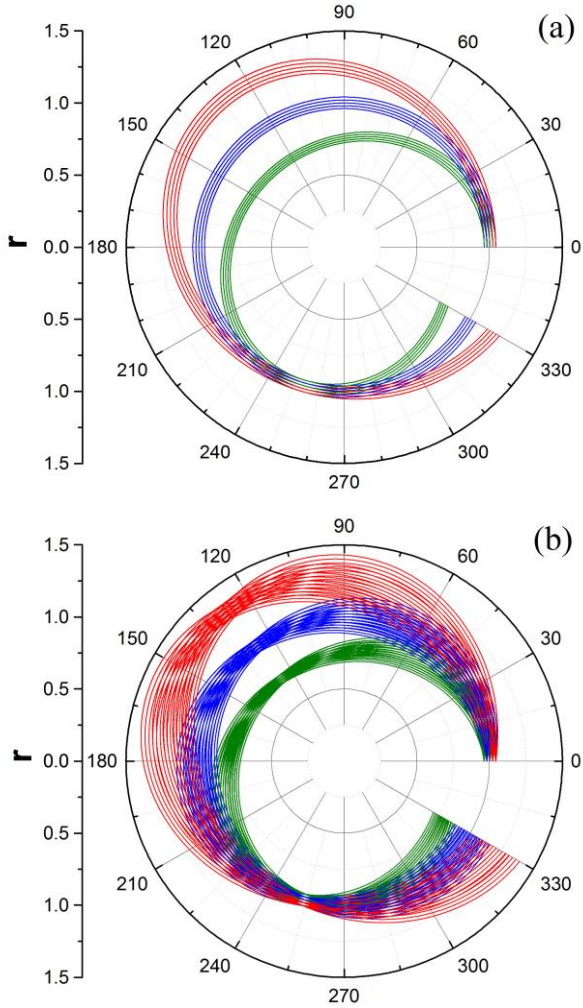

FIG. 3. Blurring of the point of fall at $\pi/\sqrt{2}$ caused by the finite entering slit and the tilt of incident angle depending on the relative energy $\lambda$. (a) Only the trajectories of normal incident ions are plotted. (b) Including the tilt of incident angle. The divergence of starting radius between two neighboring trajectories is 2%, the divergence of incident angle between two neighboring trajectories is 4°, and the energy of each bunch is 0.7, 1.0 and 1.3 time of the circular orbit energy, respectively.

First, as illustrated both in Fig. 1(a) and Fig. 1(b), the calculation confirms that the ions starting from a same point will be refocused to certain radiuses depending on their energies after revolving an angle of $\pi/\sqrt{2}$, either their kinetic energy is nearby or is far from the circular orbit energy. In order to deduce the relationship between the point of fall at $\pi/\sqrt{2}$ versus the energy, let us consider an ion with energy $E = \lambda E_c = \lambda q\kappa/2$, starting from $r_0$ and perpendicular to the radius. Note that after revolving an angle of $\pi/\sqrt{2}$, its momentum will be perpendicular to the radius again, according to the period of $\sqrt{2}\pi$ of $r(\theta)$. Denote the ratio between the radius of arrival and starting by $\eta = r_1/r_0$, where $r_1$ is the radius of the point of fall at $\pi/\sqrt{2}$, considering conservation of energy and angular momentum, we obtain

$$\lambda = 2\eta^2 \ln(\eta)/(\eta^2 - 1). \qquad (2)$$

Practically, it can be calculated from its series form

$$\lambda = \eta - \frac{1}{6}(\eta-1)^2 + \frac{1}{30}(\eta-1)^4 - \frac{1}{30}(\eta-1)^5 + O(\eta-1)^6, \qquad (3)$$

and an inverse relation can be represented as

$$\eta = 1 + (\lambda-1) + \frac{1}{6}(\lambda-1)^2 + \frac{1}{18}(\lambda-1)^3 \\ - \frac{11}{1080}(\lambda-1)^4 + \frac{7}{648}(\lambda-1)^5 + O(\lambda-1)^6. \qquad (4)$$

Second, as illustrated in Fig. 1(a), the calculation shows that if the ions start from a same point with same angular momentum, they will be almost perfectly refocused to the same radius after revolving an angle of $\pi/\sqrt{2}$. As a consequence, it reveals that the refocusing of the ions with same energies is inexactitude, which is only valid for small divergence of the incident direction, as shown in Fig. 1(b). The blurring of the point of fall at $\pi/\sqrt{2}$ caused by the tilt of incident angle is studied for trajectories both concerning to the circular orbit energy and other energies. It is well known that when the direction of an incoming ion with the circular orbit energy deviates from the normal direction, either left or right, its will fall inside of the circular orbit.[6] This is still true for other energies, which can be seen in Fig. 1(b). Let $r_1'$ denote the radius of the point of fall at $\pi/\sqrt{2}$ concerning to the tilt angle $\alpha$, and $\varepsilon = (r_1' - r_1)/r_0$ represent the blurring shift of the point of fall. The dependence of $\varepsilon$ on $\alpha$ and $\lambda$ is demonstrated in Fig. 2. An useful formula has been summarized from the calculations to roughly estimate the blurring in designs,

$$\varepsilon \approx -1.4\lambda^{1.63}\alpha^2, \tag{5}$$

in which the tilt angle $\alpha$ is in unit of radians. The error is typically within 10% when $0.6 < \lambda < 1.4$ and $\alpha$ is within $\pm 10°$.

Third, the finite size of the entering slit will also blur the point of fall at $\pi/\sqrt{2}$. In consideration of the re-focusing nature mentioned above, we only take into account those normal incident ions as a typical representation, shown in Fig. 3(a). Let $\delta r_0$ and $\delta r_1$ denotes the width of the entering slit and the spread width the point of fall at $\pi/\sqrt{2}$, respectively. Direct calculation gives

$$\delta r_1 / \delta r_0 = \eta, \tag{6}$$

when taking consider of the definition of $\eta = r_1/r_0$, where $r_0$ and $r_1$ is the radius of the starting and the falling point, respectively.

In practice, both the size of the entering aperture and the opening angle of the source area to the entering aperture should be taken into account, as demonstrated in Fig. 3(b). To ensure the energy resolution, a practical criterion is chosen as the blurring of the point of fall at $\pi/\sqrt{2}$ caused by both factors are comparable to the position resolution of the detector $\Delta x$,

$$\eta \delta r_0 \sim 1.4\lambda^{1.63}\alpha^2 r_0 \sim \Delta x. \tag{7}$$

### III. REALIZATION

The realization of the spectrometer is sketched in Fig. 4(a). A pair of ring electrodes are employed to build the framework of the spectrometer. The electrodes are made of stainless steel, and assembled by a pair of circular slots which are made of teflon. The radius of the inner electrode surface is $r_i = 50\text{mm}$, and the outer one is $r_o = 76\text{mm}$, therefore a gap of 26 mm between the two electrodes is created. Both electrodes have heights of 60 mm and thicknesses of 3 mm.

A grounded skimmer tube was employed to collimate and to guide the ions into the electrostatic field. Its length is 115 mm, with an inner and an outer diameters of Φ3 mm and Φ6 mm, respectively. At its both ends there are changeable skimmers with apertures of Φ0.2 mm, Φ0.5 mm, Φ1 mm and Φ3 mm, with tapered shape for the sake of minimizing the field distortion. Its exit aperture is adjusted to the center between the two ring electrodes, and its axis is aligned to be perpendicular to the radius. Therefore, the center circular orbit has a radius of $r_0 = 63\text{mm}$ in the present layout. Alignment holes are presented both to align the skimmer tube as well as to adjust the spectrometer in order to aim at the collision area on the target. The angles among the incident HCI beam, the solid surface and the skimmer tube can be changed.

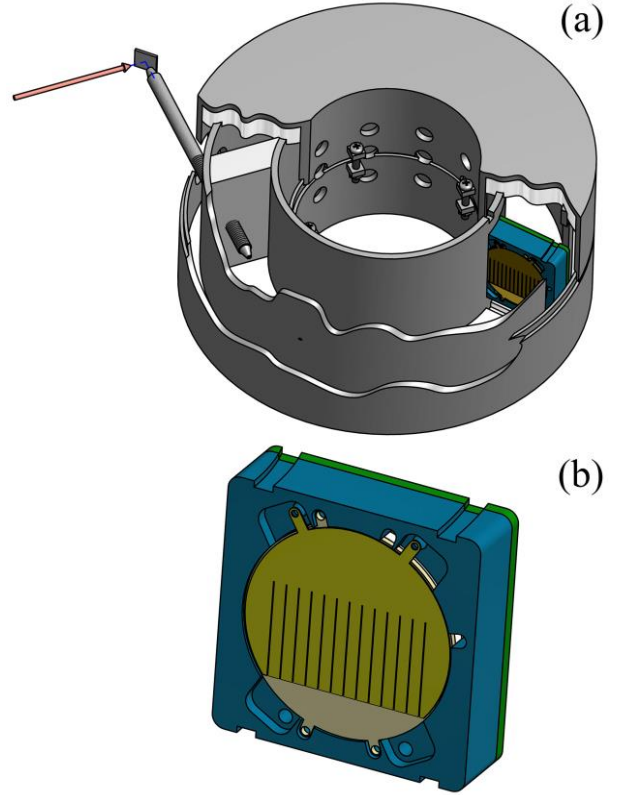

FIG. 4. The schematic view of the 127° cylindrical energy spectrometer. (a) The spectrometer. The radius of the inner and the outer electrode is 50 mm and 76 mm, respectively. A skimmer tube is employed to collimate and to guide the ions into the electrostatic field, and a position sensitive MCP detector is equipped to distinguish different orbits. (b) The detector. A comb-shaped shelter is employed to ensure the absolute positions of the point of fall and to regularize the absolute receiving area.

In order to ensure the absolute positions of the point of fall, a brass comb-shaped shelter, as shown in Fig. 4(b), was fixed at 127.3° downstream of the outlet of the skimmer tube. Its thickness is 0.1 mm, with 13 slits of 0.2 mm width each, and a period of 2 mm. A position-sensitive micro-channel plate (MCP) detector is placed behind the shelter. Therefore, we always obtain 13 peaks on the position spectra, and for each peak its absolute position is known and only its area is interested. We note that the detector surface is biased to -3kV to post-accelerate the low energy ions, in order to increase its detecting efficiency. The assembly is shielded by a grounded shell made of stainless steel to restrain the severe background, which we have suffered in earlier experiments stemmed from direct ion-solid interactions. All the materials of the spectrometer are nonmagnetic to avoid unexpected deflection.

The effective position resolution of the MCP detector is 0.2 mm depending on to the comb-shaped shelter in the present design. According to Eq. (7) and considering the size of the collision area on the target, the inlet and the outlet skimmers are selected to be with Φ3 mm and Φ0.2 mm apertures, respectively.

If the working voltage between the outer and the inner electrodes is denoted by $U = U_o - U_i$, the corresponding parameter $k$ of electrostatic field is

$$k = U/\ln(r_o/r_i). \quad (8)$$

Therefore, the circular orbit energy in the present design is

$$E_c = qU/2\ln(r_o/r_i) \approx 1.194qU. \quad (9)$$

In order to further minimize the field distortion induced by the grounded skimmer tube, we need to ensure that the center circular orbit, i.e., the r=63mm circular orbit, is always at the ground potential. Therefore, the inner and the outer electrodes are biased with certain negative and positive voltages, respectively.

$$U_i = U\ln(r_i/r_c)/\ln(r_o/r_i) \approx -0.552U, \quad (10a)$$

$$U_o = U\ln(r_o/r_c)/\ln(r_o/r_i) \approx +0.448U. \quad (10b)$$

At last, we need to transform experimental position distributions $D_i(r)$ to energy spectra $Y_i(E)$ at each working voltage $U_i$, and subsequently joint these energy spectra to obtain the final complete one. The key point of the transform is that a same segment $\Delta r$ in the position spectra corresponds to a different segment $\Delta E$ in the energy spectra, according to different $r$ and different working voltages. At each working voltage, transform from $D_i(\eta)$ to $Y_i(\lambda)$ should be applied at the first step. In consideration of the conservation of particle number holds for *any* corresponding integration range

$$\int_{\eta_1}^{\eta_2} D_i(\eta)\mathrm{d}\eta \equiv \int_{\lambda_1}^{\lambda_2} Y_i(\lambda)\mathrm{d}\lambda, \quad (11)$$

directly deducing from Eq.(2) one obtains

$$\frac{D_i(\eta)}{Y_i(\lambda)} = \frac{\mathrm{d}\lambda}{\mathrm{d}\eta} = \frac{2\eta[\eta^2 - 1 - 2\ln(\eta)]}{(\eta^2 - 1)^2}. \quad (12)$$

At the second step, relative ratios among $Y_i(\lambda)$ should be considered. A same segment $\Delta r$ in position spectra always corresponding a same $\Delta\lambda$. Taking in account that $\Delta E = \Delta\lambda E_c = \Delta\lambda qU_i/2\ln(r_o/r_i)$, then obviously $Y_i(\lambda)$ should be divided by an additional factor of $U_i$ to obtain right ratios in the final complete energy spectrum. For example, the area ratios among peaks in the position spectrum obtained by the present design should be divided by the factors of $1.062U_i$, $1.052U_i$, $1.042U_i$, $1.032U_i$, $1.021U_i$, $1.011U_i$, $U_i$, $0.9894U_i$, $0.9789U_i$, $0.9684U_i$, $0.9579U_i$, $0.9475U_i$ and $0.9372U_i$ to obtain the ratios in the final energy spectrum.

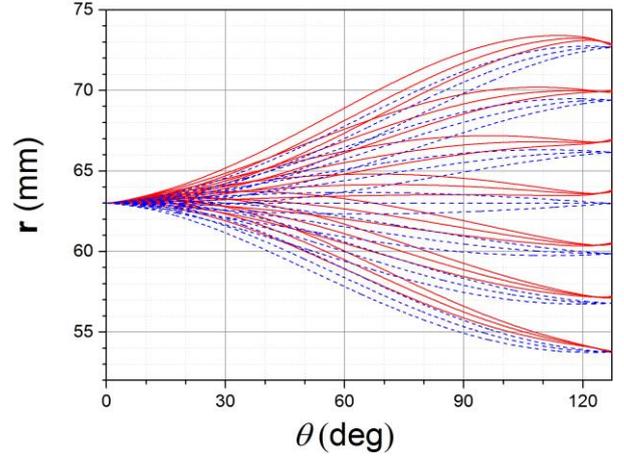

FIG.5. Comparison between the ionic trajectories in the actual field (i.e., including fringing fields, the red solid curves) and in the ideal field (the blue dashed curves).

The SIMION software[27] is employed to evaluate the effects of fringing fields. A comparison between ionic trajectories in the ideal field and in the actual field including the fringing parts of both the grounded skimmer tube and the grounded receiving surface of the detector is illustrated in Fig. 5. It shows that the trajectories in the actual field are systematic outside comparing to the ideal field, and the deviations of point of falls between these two cases are within 1 mm. The simulations show that the most important effect of the fringing fields comes from the skimmer tube, by weakening the radial field nearing its outlet. It also shows that a thinner skimmer tube with a sharper outlet result in a weaker fringing effect. In the present design the trajectories are much less affected than the case of Oshima et al.[21,22], in which the whole starting surface is grounded. It also confirms that the optimal refocusing angle is slightly smaller than $\pi/\sqrt{2}$ when the receiving surface is grounded, as pointed out by Oshima *et al.* to be about 126.5° under their conditions.[22] Besides, the simulations also show that in the present design the height of electrodes, 60 mm, has negligible effects on the electrostatic field.

**5 / 6**

## IV. CONCLUSION

In this work, an analytical relation between the point of fall of an ion and its energy, as well as the transform from a position spectrum to an energy spectrum are presented of a multichannel 127° cylindrical energy spectrometer with an ideal field. The blurring of the point of fall caused by the divergence of the incident angle and the finite size of the entering slit are considered. A 127° cylindrical spectrometer equipped a position-sensitive MCP detector has been designed and constructed, and was employed to measure the sputtered ions in HCI with solid surface interactions.[25]

## ACKNOWLEDGMENTS

Fangfang Ruan, Mingwu Zhang, Wei Wang, Jing Chen, Caojie Shao and Rongchun Lu joined in the machine design, the commissioning or the experiments. Junliang Liu helped on the using of the SIMION software and on the drawing of Fig. 4. Yingli Xue carefully read the manuscript. Some experiments were carried out at the 320 kV platform for multidiscipline research with highly charged ions of Institute of Modern Physics, Chinese Academy of Sciences. This work is partially supported by the National Natural Science Foundation of China under Grant Nos. 11275240 and U1332206.

[1] D. Roy and D. Tremblay, Rep. Prog. Phys. **53**, 1621 (1990).

[2] C. Linsmeier, Vacuum **45**, 673-690 (1994).

[3] R. C. G. Leckey, J. Electr. Spectr. Relat. Phenom. **43**, 183-214 (1987).

[4] W. Steckelmacher, J. Phys. E: Sci. Instrum. **6**, 1061 (1973).

[5] C. C. Chang, Surf. Sci. **25**, 53-79 (1971).

[6] A. L. Hughes and V. Rojansky, Phys. Rev. **34**, 284-290 (1929).

[7] A. L. Hughes and J. H. Mcmillen, Phys. Rev. **34**, 291-295 (1929).

[8] M. Arnow and D. R. Jones, Rev. Sci. Instrum. **43**, 72-75 (1972).

[9] M. Arnow, J. Phys. E: Sci. Instrum. **9**, 372 (1976).

[10] E. Tomková, K. Kokešová and V. Nehasil, Czech J. Phys. **35**, 621-629 (1985).

[11] A. Salop, D. E. Golden and H. Nakano, Rev. Sci. Instrum. **40**, 733-735 (1969).

[12] Y. Gao, R. Reifenberger and R. M. Kramer, J. Phys. E: Sci. Instrum. **18**, 381 (1985).

[13] J. A. Stroscio and W. Ho, Rev. Sci. Instrum. **57**, 1483-1493 (1986).

[14] D. J. O. Connor, J. Phys. E: Sci. Instrum. **20**, 437 (1987).

[15] M. Nishijima, Y. Kubota, K. Kondo *et. al.*, Rev. Sci. Instrum. **58**, 307-308 (1987).

[16] V. Manoram, S. K. Pawar, S. V. Bhoraskar *et. al.*, J. Phys. D: App. Phys. **24**, 1207 (1991).

[17] T. Nagao, Y. Iizuka, M. Umeuchi *et. al.*, Rev. Sci. Instrum. **65**, 515-516 (1994).

[18] B. Hird, P. Gauthier and R. A. Armstrong, Rev. Sci. Instrum. **66**, 3273-3279 (1995).

[19] J. Rubio-Zuazo, M. Escher, M. Merkel and G. R. Castro, Rev. Sci. Instrum. **81**, 043304 (2010).

[20] Http://En.Wikipedia.Org/Wiki/Binet_Equation.

[21] C. Oshima, R. Franchy and H. Ibach, Rev. Sci. Instrum. **54**, 1042-1046 (1983).

[22] C. Oshima, R. Souda, M. Aono and Y. Ishizawa, Rev. Sci. Instrum. **56**, 227-230 (1985).

[23] R. Franchy and H. Ibach, Surf. Sci. **155**, 15-23 (1985).

[24] D. E. David, D. B. Popović, D. Antic and J. Michl, J. Chem. Phys. **121**, 10542-10550 (2004).

[25] F. Ruan, D. Yu, Z. Mingwu *et. al.*, Rev. Nucl. Phys. **26**, 227-230 (2009).

[26] J. Liu, D. Yu, F. Ruan *et. al.*, Rev. Sci. Instrum. **84**, 036107 (2013).

[27] Http://Simion.Com/.